
\hfill CITA/95/21


\newbox\abstr
\par
\def\reindent{\par\noindent\parskip=0pt\hangindent=3pc\hangafter=1}
\def\window#1{\parskip=0pt\voffset=1.2truein
\setbox\abstr=\vbox{\hsize 4.0truein{#1}}
\hbox to \hsize{\hfill\box\abstr\hfill}}
\window{\centerline{\bf  STAR FORMATION AND CHEMICAL EVOLUTION}
\centerline{\bf IN DAMPED LY$\alpha$ CLOUDS}
\vskip10pt
\reindent
\centerline{ROBERT A. MALANEY AND BRIAN CHABOYER}
\vskip10pt
\reindent
\centerline{Canadian Institute for Theoretical Astrophysics,}
\reindent
\centerline{60 St. George St., University of Toronto,}
\reindent
\centerline{Toronto, Ontario,  Canada M5S 1A7.}
\reindent
\centerline{Email: malaney@cita.utoronto.ca; chaboyer@cita.utoronto.ca}
\vskip10pt
}
\par

\vskip 1.5truein

\midinsert\narrower
\centerline{ABSTRACT}
Using the redshift evolution of the neutral hydrogen density, as inferred
from  observations of damped Ly$\alpha$ clouds, we  calculate
the evolution of star formation rates and elemental abundances in the universe.
For most observables our calculations are in rough agreement with previous
results based on the instantaneous re-cycling approximation (IRA).
However, for the key metallicity tracer Zn, we find a better match to the
observed  abundance at high redshift than that given by the
constant-yield  IRA model. We investigate whether the redshift evolution of
deuterium,  depressions in the diffuse extragalactic gamma-ray background,
and measurement of the MeV neutrino background may help determine
if observational bias due to dust obscuration is important.
We also indicate how the importance of dust on the calculations can be
significantly reduced if correlations of  the  HI column density with
metallicity are present. The possibilities for measuring $q_o$
with  observations  of elemental abundances in damped Ly$\alpha$ systems
are discussed.
\endinsert

\vfill
\noindent

\vfill
\eject

\hoffset=0 true in
\hsize=6.75 true in
\voffset=0 true in
\vsize=9 true in
\overfullrule=0pt
\abovedisplayskip=15pt
\belowdisplayskip=15pt
\abovedisplayshortskip=10pt
\belowdisplayshortskip=10pt

\def\gapp{\mathrel{\raise.3ex\hbox{$>$}\mkern-14mu
              \lower0.6ex\hbox{$\sim$}}}
\def\hbar{{\mathchar'26\kern-.5em{\it h}}}



\def\gtorder{\mathrel{\raise.3ex\hbox{$>$}\mkern-14mu
             \lower0.6ex\hbox{$\sim$}}}
\def\ltorder{\mathrel{\raise.3ex\hbox{$<$}\mkern-14mu
             \lower0.6ex\hbox{$\sim$}}}

\baselineskip=12truept plus 1pt minus 2pt

\centerline{1. INTRODUCTION}

It is widely believed that the damped Ly$\alpha$ systems contain the
vast majority of the cool neutral gas at high redshift.
 Recent observational
studies  which attempt to determine the redshift evolution  of this gas
(Pettini {\it et al.} 1994;
Lanzetta {\it et al.} 1995;  Storrie-Lombardi
{\it et al.} 1995), open up the possibility of seriously exploring
cosmic chemical evolution  in the  early universe. This is
especially true if the cool gas can be considered to be  a closed box system,
since this allows  for the most direct inference of
the  universal star formation rate (SFR). Such  a direct knowledge of
the SFR greatly reduces the input uncertainties to the chemical evolution
equations.  Based on their observations, and the use of instantaneous
re-cycling approximation  (IRA) calculations,
Lanzetta {\it et al.} (1995) give a series of compelling  arguments which argue
in favor of the closed box models, and against inflow/outflow models.
In the calculations reported here we will assume that the chemical evolution
equations are indeed accurately described by closed box models.

The reasons for the present study are three-fold. First, we wish to
carry out chemical evolution calculations
which do not utilize the IRA.
In doing so we wish to assess the reliability of previous calculations
which make this assumption. Secondly, we wish to explore the possibilities
of testing cosmology with future observations of elemental abundances
in damped Ly$\alpha$ systems. Third, we wish to explore the influence
on the chemical evolution of
observational bias introduced by dust obscuration
of background quasars, and to determine possible tests for such bias. Recently,
Pei and Fall (1995) have argued that such dust contamination could
in fact strongly influence the predicted chemical evolution.

\centerline{2. CHEMICAL EVOLUTION}

The cosmological gas density $\Omega_{g}(z)$
 is defined as the comoving
gas mass density  at a redshift  $z$ in units of the present
 critical density $\rho_c$. It
is given by

$$
\Omega_{g}(z)={H_o \mu m_h\over c \rho_c}\int^{N_{max}}_{N_{min}} f(N,z)NdN
\eqno(1)
$$
where  $c$ is light speed, $H_o$ is Hubble's constant,
$\mu$ is the mean particle mass per $m_h$ (the mass of an H atom),
$N$ is the column density of HI, and $f(N,z)dNdX$ is the number of absorption
systems per line of sight with column density in the interval $N$ to $N+dN$
and ``absorption distance'' in the interval $X$ to $X+dX$.
The coordinate
$X$ is defined through
$$
X(z)=\int^z_0(1+z) (1+2q_oz)^{-1/2}dz
\eqno(2)
$$
where $q_o$ is the de-acceleration parameter. We assume any
molecular abundances to be negligible and take
 $\Omega_g(z)=1.3\Omega_{\rm HI}(z)$.

Assuming zero cosmological constant and neglecting radiation,
the relation between redshift $z$  and time $t$ can be written
$$
{dt\over dz}=-{H_o^{-1}\over 1+z} \biggl [\Omega_m(1+z)^3 -\Omega_k(1+z)^2
\biggr ]^{-1/2}
\eqno(3)
$$
where
$$
\Omega_m=(\rho/\rho_c)_o=8\pi G \rho/3 H_o^2 ; \ \ \ \Omega_k=Kc^2/H_o^2 \ \
(K=\pm 1,0) \ \ \ .
$$
{}From the above relations the redshift evolution of the cosmic gas
in a closed box model is
described by the following equation

$$
{d\Omega_{g}\over dz}=   \biggl \{\ { \int^{m_{up}}_{m(z)} (m-m_r)
 \Psi(z^f_m) \Phi(m) dm
-\Psi(z) } \biggr \} {dt\over dz} \ \ \ .
\eqno(4)
$$
Here $z^f_m$ is the formation  redshift at which a star of mass $m$ is
returning gas back to the interstellar medium at the current redshift
 $z$, $\Psi$ is
the SFR, and $\Phi$ is the initial mass function (IMF).
The lower limit of integration $m(z)$ is the minimum  stellar mass which
can be returning gas to the interstellar medium at redshift $z$.
For the remnant mass
function  $m_r$, the relations of Iben and Tutukov (1984) are adopted;
and for the stellar lifetimes the relations of Scalo (1986) are adopted.
Note that a small correction  due to the presence of
type~Ia supernovae is actually added  to eq.~(4) in our calculation
 (see discussion below); for the
evolution of the gas this contribution is unimportant and for clarity
it is not explicitly  shown above.
We assume a power law for the IMF viz. $\Phi(m) \propto m^{-(1+x)}$,
 ($x$ is termed the slope of the IMF) and it is
 normalized through the relation
$$
\int^{m_{up}}_{m_{low}} m \Phi(m)dm=1 \ \ \ .
\eqno(5)
$$
The upper and lower mass limits are taken to be $m_{up}=40 M_\odot$ and
$m_{low}=0.08 M_\odot$, respectively. In this notation $x=1.35$ corresponds
to the Salpeter IMF.

We choose to parameterize the observed  and dust-corrected
evolution of $\Omega_{\rm HI}(z)$
through the functional forms given by Pei and Fall (1995).
The observational data
is then described through
$$
\Omega_{\rm HI}(z)={\Omega_{\rm HI}(\infty) \exp(\alpha z)\over \exp(\alpha z)+
\exp(\alpha z_b)}
\eqno(6)
$$
where $\alpha$ and $z_b$ are two of the available fitting parameters
(dependent on the adopted cosmological model). Figures~1 displays
appropriate fits to the observed data for   $q_o= 0.5$.  The solid curve
is an appropriate fit to the data of Lanzetta {\it et al.} (1995).
The dashed curve
displays a dust-corrected  evolution of the  Lanzetta {\it et al.} (1995) data
typical of that calculated by Pei and Fall (1995). Clearly, the effects of dust
can have a dramatic impact on the inferred  evolution
of the neutral hydrogen. We will return  to this effect later.
 For the time being we assume the observed evolution of
$\Omega_{\rm HI}(z)$ approximates closely  the true evolution.
The dotted curve of figure~1 is an
appropriate fit to the  preliminary
 data of Storrie-Lombardi {\it et al.} (1995),
which suggests a lower value of $\Omega_{\rm HI}$ at $z>3$.

Defining $\Omega_{i}$ as $\Omega_{g} X_i$, where $X_i$ is the  gas
mass fraction of element $i$, we can follow the evolution  of the
elemental abundances with $z$ through the following equation
$$
\eqalign {&{d\Omega_{i}\over dz}=
 \biggl [   \int^{3}_{m(z)} m \Psi(z^f_m) {\Phi(m)} Y_i(z^f_m) dm \cr
& + r \int^{16}_{m3(z)} m_b \biggl \{\int^{0.5}_{\mu_m} f(\mu)
  \Psi(z^f_m)  Y_i(z^f_m)d\mu \biggr \}  \Phi(m_b) dm_b \cr
& + (1-r) \int^{16}_{m3(z)} m \Psi(z^f_m) {\Phi(m)} Y_i(z^f_m) dm
 +\int^{m_{up}}_{m16(z)} m \Psi(z^f_m) {\Phi(m)} Y_i(z^f_m) dm
-\Psi(z)X_i \bigg ] {dt\over dz} \cr }
\eqno(7)
$$
Here $Y_i$ is the elemental stellar yield, written as  the mass of element $i$
in the ejecta divided by the initial  mass of the star
 from which it was ejected.
We have adopted the yield calculations of Renzini and Violi (1981)
 for low mass stars, of
Woosley and Weaver (1995) for massive stars,
 and of Thielemann {\it et al.} (1986)
 for type~Ia supernovae.
The parameter $r$ parameterizes the contribution of the type~Ia supernovae
to the elemental evolution.  Other than the rate of type~Ia supernovae,
the only observable  reported here
that this contribution  significantly affects is the Fe abundance.
We assume here a constant value $r=1/200$, which provides agreement with
the present extragalactic supernova rates (see later).
The additional integral here is
over the binary distribution function $f(\mu)$, where $\mu$ is the ratio
of secondary mass to total binary mass $m_b$. We have adopted the distribution
function of Greggio and Renzini (1983), $f(\mu)=24 \mu^2$.
Note that the lower limits of integration  $m3$ and $m16$
have the same meaning as before except
that they have lower limits of $3$ and $16M_\odot$, respectively.
 For most IMF's
raising the upper limit of  $m_{up}=40M_\odot$ makes
little difference to the results.

We  now solve the above equations in
order to determine the chemical evolution of the cosmic gas as a function of
redshift.

\centerline{2.1. {\it Star Formation Rates}}

We wish to compare our results with those obtained previously
using an IRA (Lanzetta {\it et al.}. 1995, Pei and Fall 1995).
Many predicted properties of the evolving gas
arise from the calculated SFR. So let us first investigate the SFR determined
from the above relations
and compare them with those determined assuming an IRA.
We determine an IRA solution by setting
$\Psi(z^f_m)\rightarrow \Psi(z)$ and $m(z)$ to some constant mass
 (say $0.85M_\odot$) in eq.~(4), giving
$$
\Psi_{IRA}(z)={1\over 1-R} {d\Omega_g\over dz} {dz\over dt}
\eqno(8)
$$
where the return fraction $R$ is
$$
R=  \int^{m_{up}}_{0.85} (m-m_r)  \Phi(m) dm \ \ \ .
\eqno(9)
$$

Figure 2 illustrates the results for three different assumed IMF's
(using the solid curve of figure~1).
The dashed curves correspond to the IRA approximation and solid curves
to the real calculations. The calculations are for  a Salpeter IMF  ($x=1.35$),
and two extreme slopes of
$x=0.35$ and $x=2.35$. This range of $x$ should encompass all reasonable
IMF's.
  It is seen that the IRA
is a good approximation except at extremely flat IMF's. These trends are
as expected since for flat IMF's  relatively more matter is being ejected at
high redshift.
\medskip
\centerline{2.2. {\it Elemental Abundances}}

 Next we calculate the elemental abundances as function of redshift.
 The form of the IMF in
our calculations
is the main effect which
alters the absolute level of most predicted elemental abundances.
We find that an IMF with $x=1.7$ matches well the solar metallicity
mass fraction   $Z=0.02$
at $z\sim 0$ . The Salpeter IMF gives $Z\sim 0.03$ at $z\sim 0$.
(note  capital Z is used for metallicity and small z for
redshift). If one wishes to normalize the calculations to solar type
 metallicities at low redshift then values of $x$ which do not
deviate far from $1.7$ are required.
 Figure~3a shows the redshift evolution of $Z$, Si, Cr, Fe and Zn
for an IMF slope of $x=1.7$ and for $q_o=0.5$.
Also shown is  the Zn abundance as  inferred by Pettini {\it et al.} (1994)
 from the weighted average of $17$ damped Ly$\alpha$ ZnII
observations.
A complete compilation of the presently available abundance data can be found
in Timmes {\it et al.} (1995a). Here we focus on the Zn abundance since
it is widely used as the metallicity tracer in damped Ly$\alpha$
systems. This is largely due to the fact that
its depletion onto dust grains is anticipated to be
small. On the other hand, Cr is anticipated to be highly depleted
onto dust grains, and its abundance probes the nature of the dust
rather than  the metallicity (see Pettini {\it et al.} 1994 for further
discussion).

The predicted  evolution of the elemental abundances
 are in broad agreement with IRA calculations. However, the
predicted Zn curve
  is significantly lower
relative to  a constant-yield IRA calculation for this element.
This effect is a result of  the metallicity dependence  (i.e redshift)
dependence of
the Zn yields from the type~II supernovae
 calculations. To illustrate this point,  the dashed curve
of figure 3a is calculated
assuming  a constant $Y_{Zn}$ yield at all redshift. Although a mass
dependence is still included here, the  only yields  utilized are
from the solar-metallicity type~II  calculations of Woosley and Weaver's (1995)
tabulations.
It can be seen that the inclusion of metallicity dependent yields
results in a lower predicted Zn abundance.  As Zn is
the principal metallicity tracer in damped Ly$\alpha$ clouds, this effect
helps to alleviate the apparent contradiction of the cosmic G-Dwarf problem
as discussed by Lanzetta {\it et al.} (1995). The basic point is that direct
use of Zn as a metallicity tracer at high redshift, tends to underestimate
the true total metallicity Z  at that redshift
(by factors $\sim 3 \ {\rm to}\  7$).
 The other elemental abundances shown are  only modestly
affected  by  the metallicity dependence of the stellar yields (at the level of
 $\sim 25\%$).

 In their study of galactic chemical evolution,
Timmes {\it et al.} (1995b) note  a similar effect of the Zn yield dependence
on metallicity with regard to
  the  galactic Zn/Fe evolution with Fe.  Their calculations
are in general agreement with
stellar observations of Zn   in the galaxy.
However, it should be noted
that although the galactic Zn/Fe ratios
have scatter at a level roughly consistent with the metallicity dependent
yield,
 there is
no obvious increasing trend of the  observed
Zn/Fe ratio with Fe/H (Sneden {\it et al.} 1991).
 At the present time, it is not clear what the source
of this potential discrepancy could be. Perhaps it is a peculiarity of the
chemical evolution of our own galaxy, perhaps
 some correction (such as non-LTE effects) is required
for the inferred stellar abundances, or perhaps
the calculated stellar yields for Zn require adjustment.
Of course if   the latter  is correct, then
the previous discussion is modified, and constant-yield IRA
calculations may in fact be reliable for the cosmological Zn abundance.

To put the above discussion another way; if one reads
 from the  available stellar Zn/Fe data that
a Zn underabundance at high redshift equals the underabundance of
any other metal, then the dashed curve of figure~3a is the more
appropriate Zn curve to use. However, this must mean that the
 stellar yield tabulations for Zn in massive stars (the principal source
of Zn) we have utilized here
are in error. There  are of course potential sources of error
 in the yield tabulations,
such as mass loss (which we neglect for massive stars),
 nuclear physics, and the kinetic energy of type-II supernovae
(we have adopted here the high kinetic energy yields  of Woosley and Weaver).
Other more minor issues which relate to the stellar yields are
the low mass cut-off
of $11M_{\odot}$ we adopt, and the fact that  the distribution of
 elements making up
the total $Z$ will not correspond directly with the tabulations.
 Future theoretical and observational work should
 finally resolve the issue of the Zn evolution. At the present
time we  simply make the point that
use of the  current state-of-the-art stellar yields results in a lower
Zn abundance in the cosmic gas,
and apparently a better fit to the inferred abundance of Zn at $z\approx 2.2$.

For comparison, we show in figure~3b the evolution of the abundances using
a Salpeter IMF. Here we can see the impact that a change in slope of the IMF
can
have on the predicted abundances. As mentioned above,
 we see  how
higher metallicities are predicted by the shallower slope. Note also
the  different shape of the Zn curve, which can be attributed to the
higher metallicity at a given redshift and the consequent change in the
stellar yields at that redshift. With regard to the discussion above,
 this implies that the accuracy of
Zn as a metallicity tracer is  also  dependent on the IMF.

We also investigated the effects of using the  dotted curve of figure~1.
This curve approximates the data
of Storrie-Lombardi {\it et al.} (1995), which indicates lower gas densities
at high redshift. However, we find no large effect on the predicted abundances
with the use of this different $\Omega_{\rm HI}(z)$ curve.
For example, we find the abundances to be
within $\sim 0.4$ dex of the abundances shown in figure~3a.

The calculations above  were repeated for $q_o=0$. The
differences in the elemental abundances for this modified cosmology
were also found to be quite small. At $z=3$ they were roughly 0.3 dex lower
than the $q_o=0.5$ calculation, with  differences
 slightly smaller (larger) than this at
lower (higher) redshift. This leads us to the next topic.

\medskip
\centerline{2.3. {\it Testing Cosmology}}

In principal, one can utilize the abundance history in the
damped Ly$\alpha$ systems to probe cosmology. Indeed,
recently Timmes {\it et al.} (1995a) have advocated such a program,
by utilizing the galactic age-metallicity relation to scale
galactic chemical evolution directly into redshift space.
This assumes that the chemical evolution history of the galaxy
is identical to that in all  of the observed damped Ly$\alpha$ systems.
We prefer to  utilize  calculations based on
 the observed evolution of the cosmic
 neutral gas  as inferred from damped Ly$\alpha$ systems
to address this
issue.

One can test the cosmological model by looking
at the abundance ratio of two elements, where one of the elements is
produced mainly by type~II supernovae (eg O), and the other is produced mainly
by type~Ia supernovae (eg Fe).
 Basically one is trying to use the
abundances to test the relationship between time  and redshift, where
the ``clock'' utilized is the lifetime of type~Ia
precursors. Since these supernovae are
dependent on the evolution of relatively low mass systems, enough time
must have passed ($\sim 1$Gyr) before they can be produced in
significant numbers.
Once the  number of type~Ia's  become  significant the O/Fe ratio
 should start to decline.
 This idea has been advocated before by
Hamann and Ferland (1993) in the context of the evolution of
QSO broad line gas.

Here we use our calculations to investigate the possibilities of
measuring $q_o$ by measuring abundances in damped Ly$\alpha$ systems
(the IRA model would be of no value in this regard).
Figure~4 shows the O/Fe (by mass) as function of the redshift.
The solid curve is
for $q_o=0.5$, and dashed curve is
for $q_o=0$.
 The effect of the cosmology on the abundance ratio is
 clearly seen here.
It is unfortunate that the type~Ia supernovae rate exhibits a
  smooth  rise -- a   more  abrupt turn-on of the type~Ia's
 would  result in
a more distinctive drop in the O/Fe ratio.
In order to use these curves as a test of $q_o$, accuracy in the abundance
ratio to better than $30\%$ will be required. It will therefore be
a  difficult experiment to
determine $q_o$  from future observations of the abundances
in the damped Ly$\alpha$ systems. A similar conclusion  is drawn regarding
determination of the cosmological constant by this means. For comparison,
the effect of varying the IMF on the O/Fe ratio is given (dotted curve).
It can be seen that for a Salpeter IMF the effect is smaller
 than the variations induced by the cosmology.

\medskip
\centerline{2.4. {\it Dust Corrected Yields}}

As mentioned earlier, the  inferred redshift
evolution  of the neutral hydrogen
 can be dramatically influenced if
observational bias introduced by dust contamination is a serious issue.
The dashed curve of figure~1 is typical of the correction that may have to
be applied to the observed (solid curve) $\Omega_{\rm HI}(z)$ evolution
(Pei and Fall 1995). To show the potential effect of dust corrections on the
elemental  abundance evolution we utilize the dust-corrected curve of
figure~1  to re-calculate the element evolution
for $q_o=0.5$. {\it If} this curve did in fact
represent the true evolution of the cosmic gas, the
evolution of the elemental abundances would be as shown in figure~5.
 When compared to the calculations of figure~3a it can
be seen that the redshift evolution of the elements are significantly affected.
These calculations can be compared directly with the closed-box
IRA calculations of
Pei and Fall (1995). Again we find that the
IRA determination of
the abundance evolution is in broad agreement with the  more exact treatment,
except for the  evolution of the Zn abundance.
Here the predicted Zn abundance seems somewhat lower and does
not provide as good a fit to the inferred abundance of Zn as
the dust-free calculation.

Clearly the potential impact of dust-corrections on the chemical evolution
of the cosmic gas can be very significant.  There is then the danger that
the ability to use observed abundances in damped Ly$\alpha$ systems
as a  probe of evolution in the
 early universe, can become seriously compromised. Not
only  could it be that a large fraction of the neutral gas is hidden, but that
the true  cosmic elemental abundances  (obtained from
a weighted average over all column densities)
are also unmeasurable. This would render the confrontation of calculation with
observation untenable.
 In view of this, we wish to turn
to  other possible future experiments which may shed some light on whether
dust contamination is   a source  of significant observational bias.

\medskip
\centerline{3. TESTS OF DUST CONTAMINATION}

One way of investigating  possible dust contamination
 is to pose the question; what other observables
are there of the  evolution of the cool neutral gas
at high redshift?
In this section we will address this question by comparing
calculations which utilize the
the observed (solid) and
dust-corrected (dashed) evolutionary   curves of figure~1?
Use of other observed or dust-corrected
 curves for the neutral hydrogen evolution will not be further studied.
In this section we are only interested in possible signals of
the  dust effects, and the two curves of figure~1 we adopt
 provide good examples by which to carry out comparison tests.
However, we do note that use of the dotted curve of figure~1
(and its similarly dust-corrected curve)
 would  only modestly impact the results below.
Also, unless otherwise stated we will adopt
$x=1.7$  since this IMF matches solar metallicities at low $z$,
and we adopt the conventional wisdom that the universe is flat
($q_o=0.5$).
  We will note circumstances where
 the proposed tests are
 significantly affected by relaxation of these  two choices.

\medskip
\centerline{3.1. {\it Deuterium Evolution}}

Deuterium is the best isotope to study in order to investigate dust effects.
The reasons for this are the following.
(a) We can assume deuterium is destroyed in stellar interiors and
never present in any significant quantities in stellar ejecta --
 this makes our calculation
independent of any  stellar yield calculation.
(b) Unless an unlikely combination
of chemical evolution effects in individual absorption clouds conspire,
a  plateau   in the  deuterium abundance $vs$. redshift plot
will indicate the primordial deuterium abundance. This
will make the  lack of an abundance determination in the
dust-hidden clouds unimportant.
(c)  Assuming the validity of
standard big bang nucleosynthesis, the primordial abundance of
deuterium is known {\it a priori} to be
confined within a narrow range. Although not as important as (a) and (b),
this last point does allow for a consistency check.

Adopting point (a) above, and
defining $\Omega_D=\Omega_{g}(z) X_D$, where $X_D$ is the gas mass fraction
of deuterium, the evolution of the deuterium  abundance can be described
by
$$
{d\Omega_{D}\over dz}=  {{H_o^{-1}}
\Psi(z) X_D  \over (1+z) [\Omega_m(1+z)^3 -\Omega_k(1+z)^2]^{1/2}} \ \ \ .
\eqno(10)
$$
The ratio of deuterium to its primordial value
can  then be calculated for the different  curves of the
neutral hydrogen evolution.
The resultant evolution of the deuterium as a function of
redshift can be seen in figure~6. It is clear that the dust-corrected
evolution of deuterium is significantly different from that calculated
using the observed $\Omega_{\rm HI}(z)$ curve. The presence of a
longer plateau in  the dust-corrected calculation
 arises from the later onset of significant star formation in this model.
 Calculations for different IMF's
 using the observed $\Omega_{\rm HI}(z)$ curve were also investigated.
For example, the deuterium evolution for a Salpeter IMF (no dust correction)
is also shown in figure~6.
In addition, we note that IMF's with
values of $x>2$ result in little depletion of deuterium, and
can mimic the dust-corrected calculation. Clearly then, the IMF
can have an important impact on the deuterium evolution.
With regard to probing dust contamination, it is therefore
important  that the actual IMF be tied down from
other considerations such as the metallicity  production
associated with it.  For example,
IMF's with $x>2$ result  in very low metallicity at $z=0$. Changes
in the cosmology had smaller effects  on the deuterium evolution
than those imposed by  dust effects.

One can see that in principal, future observations of deuterium in
damped Ly$\alpha$ systems can be a probe for the presence of dust-induced
observational bias. However, the accuracy required to distinguish between
the observed and dust-corrected
curves is $\sim 10\%$, thus making this test extremely difficult in practice.
As an aside, it is interesting to note that the calculations predict the
present  cosmic
deuterium abundance to be reduced by only $30 - 40\%$ from its primordial
abundance. Figure~6 also indicates
 the redshift at which deuterium must be observed
in order to probe its true primordial abundance.

\medskip
\centerline{3.2. {\it  Extragalactic Gamma-Rays}}

The rate of type~II
supernovae,  $R_{II}$ (number/Gyr/unit comoving volume),
 can be determined through
$$
R_{II}=\rho_c
  \int^{m_{up}}_{m16(z)}  \Psi(z^f_m) {\Phi(m)} dm
+(1-r) \rho_c\int^{m_{up}}_{m11(z)}  \Psi(z^f_m) {\Phi(m)} dm
\eqno(11)
$$
where the lower limits of integration  $m16$ and $m11$
have seem meaning as before, and
 have lower limits of $16$ and $11 M_\odot$, respectively.
The rate  of  type~Ia supernovae  is given by
$$
R_{I}=r \rho_c
  \int^{16}_{m3(z)}{\Phi(m_b)}
 \biggl \{ \int_{{\mu_m}}^{0.5}  \Psi(z^f_m)  f(\mu)  d\mu \biggr \}  dm_b
\eqno(12)
$$
With $r=1/200$,
 the ratio $(R_I/R_{II})_o$ is in agreement
the  observed extragalactic supernova rates (Vandenbergh 1991).
In the context of the present study we believe this is the best way to
normalize the type~Ia contribution to the element evolution, rather than
use of a solar abundance ratio like O/Fe. Given the  large
uncertainty in the
extragalactic supernova rates, however,
 there is certainly room to alter the value
of $r$ slightly, thereby affecting the predicted Fe abundance.

The rate of type~II supernovae is shown in figure~7 for the
 different evolutionary curves. It is evident that
these rates peak at significantly different values of redshift. Due
to the later star-formation in the dust-corrected calculation, the
value of $R_{II}$ peaks at the relatively lower redshift $z\sim 2$.
This opens up the possibility of probing the importance of  dust effects
by scrutinizing the diffuse extragalactic gamma-ray background for the
presence of spectral features.

Extragalactic gamma-rays can be produced through cosmic ray
interaction processes such as
${\rm p}+{\rm p}\rightarrow \pi^o + {\rm anything} \rightarrow 2\gamma$.
A flux of gamma-rays produced in this fashion would have a negligible
contribution below  half the rest mass of the pion, $70$~MeV.
In principle,
 by assuming that the
extragalactic cosmic ray flux is directly proportional $R_{II}$,
adopting some efficiency for converting supernova kinetic energy
into cosmic-ray energy (typically $1\%$),
and knowing the  gamma-ray emissivity per atom induced by cosmic rays
 (Dermer 1986),
one  could calculate the number of
gamma-rays  as a function
of redshift. The absolute value arising from such a calculation
would depend on the normalization of $R_{II}$ at some  redshift,
and
 the shape and evolution of the cosmic-ray spectrum.
However,  the main point we wish to make here is
 that one
would anticipate the gamma-ray flux
produced by cosmic-ray interactions to peak at the redshift $z_p$ corresponding
to the peak of $R_{II}$. This being the case,  a depression in
 the diffuse extragalactic
gamma-ray background maximized at $70/(1+z_p)$~MeV should be present.
The gamma-ray
depression should be maximized at approximately $14$~MeV
if the observed $\Omega_{\rm HI}(z)$ curve represents the true evolution, and
$23$~MeV if
the dust-corrected $\Omega_{\rm HI}(z)$ curve  represents the true evolution
of the cosmic gas. Note that in this case a change in the IMF
does not alter the value of $z_p$, although it could alter the depth
of any spectral feature (smaller IMF slopes providing larger effects).
The dotted curve of figure~7, which corresponds to
the observed evolution with a Salpeter IMF,
illustrates this point.

The detectability of such a feature in the
 diffuse extragalactic gamma-ray background depends largely on the
 contribution of the cosmic-ray processes relative to the other sources of
extragalactic gamma-rays (eg AGN).  It is worth noting in this respect, that
although cosmic-ray processes are thought to dominate the
diffuse galactic gamma-ray background at $E>200$~MeV, the anticipated
 $70$~MeV spectral feature
in the diffuse galactic gamma-ray background is likely dwarfed
by brehmsstrahlung processes (Weber {\it et al.} 1980).
However, based on  galactic
cosmic-ray production of beryllium, Silk and Schramm (1992)
argue that up to $50\%$ of the diffuse extragalactic gamma-ray background
could arise from cosmic-ray processes, and that a $70/(1+z)$~MeV  spectral
feature could be detectable
 (we   caution that this latter  prediction is
based on a galactic normalization of the Be abundance and
therefore  somewhat uncertain).

There is no evidence  for  a depression in the
 diffuse extragalactic gamma-ray background in the
presently available data
(Thompson and Fichtel 1982; Gehrels and Cheung 1995).  Instrumentation on
board the
 {\it GRO}, has a sensitivity to diffuse
emission of $10^{-5} {\rm cm}^{-2}  {\rm s}^{-1}$ and an energy
resolution of $\sim 10\%$. This means that if no spectral feature is
detected by GRO then it must be
substantially below the $50\%$ level. Nonetheless,
given the above discussion, future scrutiny of the observed
diffuse extragalactic gamma-ray background in the $10-70$~MeV range
 is warranted. With future increases in sensitivity it may be possible
to probe the evolution of the neutral gas at high redshift
by this means.

\medskip
\centerline{3.3. {\it  MeV Neutrino Background }}

Knowledge of the type~II supernovae rate can also be used to
calculate the   energy  density of the
 MeV neutrino background. This neutrino background arises from
the convolved neutrino emission of all
relic  type~II supernovae .
In principal, this is a much ``cleaner''
signal of the peak in the type~II supernovae rate compared to the
gamma-ray background test, since the latter requires a series of
 assumptions  regarding conversion of supernovae kinetic energy into
cosmic ray energy. Of course, the difficulty lies in the detection of
the relic MeV neutrino background. However, we note that this background is in
fact a target of the Superkamiokande  neutrino
detector currently under construction in Japan.

 One can show that the energy density of the MeV neutrino background is given
by
$$
\rho_\nu={F_E\over H_o}\int^{z_c}_0 {n(z)dz\over (1+z)^5
 [\Omega_m (1+z)^3-\Omega_k (1+z)^2]^{1/2}}
\eqno(13)
$$
where $F_E$ is the energy per unit time emitted in neutrinos by a type~II
supernova, and $n(z)$ is the number density of type~II's.
Coupling the supernova rates shown in figure~7 with eq.(13), we calculate
that the mean energy of the MeV neutrino background assuming
$R_{II}$ as determined from the observed $\Omega_{\rm HI}(z)$ evolution
 is $\sim 10$  times that calculated
assuming a  constant comoving number density
 $n(z)=n_o(1+z)^3$, where $n_o$ is the
present type~II rate (for the constant rate we assume a cutoff at $z_c=7$).
For $R_{II}$ as determined from
 the dust-corrected $\Omega_{\rm HI}(z)$ evolution,
 we find the mean energy to be $\sim 7$ times  that
calculated with the constant rate. (Note that due to the different
supernova rates at low $z$ in figure~7, the constant rate backgrounds
are different for the two curves.) With $z_c=\infty $ one finds  the MeV
 neutrino background  calculated using the dust corrected $\Omega_{\rm HI}(z)$
evolution
 is $\sim 1.6$ that calculated using the observed
$\Omega_{\rm HI}(z)$ evolution. Indeed,
this is the main point that we wish to make here.
 Detailed calculations of the MeV neutrino
 background  determined directly from the inferred evolution of the
 neutral gas at high redshift, provides the possibility of
probing  dust contamination effects. The actual  size of the
change in  energy density
will of course depend on the assumed dust contamination.

We should caution that use
 of the MeV neutrino background as outlined above
 requires accurate
knowledge of the IMF and the cosmological model parameters.
 For example, use of a Salpeter IMF (see dotted curve of figure~7)
results in a factor $2.7$ increase in the  neutrino energy density
compared to the $x=1.7$  calculation.
 Also, the above calculations adopted $q_o=0.5$.
For comparison, calculations with $q_o=0$ show the neutrino energy
density to be lowered by a factor of $\sim 3$. The implies
that  only if the IMF  and cosmology are known can
the MeV neutrino background be used as a probe of dust. Alternatively
of course,
 one can be turn this argument around. That is,
 assuming  a reliable inference of the $\Omega_{\rm HI}(z)$ evolution
and IMF,
this technique can be used  as a measurement of $q_o$.

Due to these effects it may be worth  exploring
 in greater detail the  connection between
neutrino astronomy  and observations of damped Ly$\alpha$ systems.
Assuming a reliable  normalization of the calculations, and  the
convolution of the inferred $\Omega_{\rm HI}(z)$ evolution with the neutrino
energy spectra emitted by type~II supernovae, one could determine
absolute values of the relic neutrino flux anticipated from a given
cosmological
model.
 Further discussion  and other uses of the
MeV neutrino  background  can be found in Bisnovatyi-Kogan and Seidov
(1984); Krauss {\it et al.} (1984);  Woosley {\it et al.} 1986;
Totani {\it et al.} (1995).
This latter paper also discusses in detail the expected
detection rate of the MeV background by Superkamiokande, which is is
typically a few events per year. Actual use of this test will
likely require next generation neutrino detectors.

\medskip
\centerline{3.4. {\it  Modified Dust Correction }}

Finally, we wish to conclude our calculations with a brief note
regarding the possible magnitude of dust obscuration.
In probing the possible effects of dust obscuration, Pei and Fall (1995)
 assumed a constant dust-to-gas ratio for all column
densities at a given redshift -- a reasonable
approximation given the aims of their study. However,
 it is very unlikely that all clouds at
a given redshift possess the same metallicity, and therefore
 considerations
of a non-constant dust-to-gas ratio  at a given redshift
are of interest. Here we expand on the analysis of Pei and Fall
to further investigate this.

 For illustration, consider the
closed-box IRA result which implies that $U$
the ratio of the column density $N$ to the initial unevolved
column density $N_u$ evolves like $U=\exp(-Z/y)$. Here  $y$
is the {\it net} yield (see Tinsley 1980).
Now assuming that the metallicity is proportional to the dust-to-gas ratio $k$,
i.e. $Z=ak$ where $a$ is some constant, one can show  that if each
$N$ has a {\it unique} value of $k$ then a straightforward
extension of Fall and Pei's (1993) analysis applies.
The ``true'' and observed functions $f(N,z)$ (see eq.~1) are related through
$$
f_t(N,z)=f_o(N,z) \exp\biggl [-\beta {y\over a} \ln U
 {N_u\over 10^{21}{\rm cm}^{-2}}\zeta\biggl ({\lambda_B\over 1+z} \biggr )
\exp(-a k/y) \biggr ] \ \ \ .
\eqno(14)
$$
Here $\beta$ is related to the exponent of the observed luminosity function
of quasars, and $\zeta(\lambda)$ is the ratio of the extinction at a wavelength
$\lambda$ relative to that in the B band.
Adopting $\beta=2$ and $\zeta=(\lambda/\lambda_B)^{-1}$, figure~8
shows how the dust-correction can be substantially reduced by
this particular dependence of $k$ on $N$ (dotted curve).
We have illustrated this effect
by assuming  for {\it all} clouds $N_u=10^{22}{\rm cm}^{-2}$,
 and have normalized to
 $Z=0.002$  and  $k=0.05$ at $U=0.1$.
 The solid curve of figure~8 is that corresponding
 to the observed distribution at $z=2.5$, and the dashed curve
shows the dust correction  applied to $f_o$ assuming a constant $k=0.05$
(which is typical of the $k$ expected at $z=2.5$).
 Although
the effect of the dust in this new calculation is largely dependent on
the normalization conditions, the point remains that for a
reasonable choice of parameters the impact of dust effects can be
significantly reduced relative to a calculation assuming a constant
dust-to-gas ratio. For the calculation illustrated here the
dashed curve of figure~8 gives an $\Omega_g(z=2.5)$ value of
roughly $1.4$ times that determined using the dotted curve.

The calculation completed here is overly simplistic in
that all the damped Ly$\alpha$ systems commenced evolution
 at some high redshift
with the same column density, and that subsequent evolution of the
column density is attributed only to star formation. Of course,
the initial spectrum of the cloud column densities will be determined
by some underlying theory of structure formation, and therefore
 in a more realistic calculation one would not expect
a {\it unique} value of $k$ for each value of $N$. However, even at a given $N$
a dispersion in the range of $k$ will have little effect on our conclusion;
that a trend in which larger values of $N$ have lower values of $k$ leads to
a smaller correction to the  observed $\Omega_{\rm HI}(z)$ curves.
Indeed, the main point we wish to make here, is that a relaxation of the
constant $k$
approximation can lead to significantly reduced dust obscuration effects, and
that a relaxation of this approximation is entirely reasonable.

\medskip
\centerline{4. DISCUSSION AND CONCLUSIONS}

Studies of chemical evolution are normally plagued by the many-parameter facet
of the calculations, and
there are many factors which can influence of the details of
 the calculations reported here. Uncertainty in the IMF,
 stellar yields, stellar lifetimes, and stellar mass loss,
can all play a role in influencing the predicted abundances.  We have further
explored this parameter space  within  what we regard to be a
 reasonable ranges of uncertainties. The changes we find in
the predicted abundances are roughly  as expected. For example,
 changes  of factor two
in the  stellar yields results in   roughly factor two changes
 in the predicted abundances.
Our basic conclusions regarding the chemical evolution of the elements
are not seriously affected by reasonable variations of the above parameters.

Another potential source of error, is the insistence that the present
metallicity from our calculations approximates the solar metallicity.
 In doing so we have assumed
that the weighted average of  the metallicity in
{\it all} galaxies at the present epoch will give roughly the solar value.
Although there is some evidence to support such a
view, one must remember that this is
a substantial assumption. It is normalization to $Z\sim 0.02$ at small
redshift that determined the slope of the IMF in most of our calculations.
We  do not believe that normalization at $z=0$
to the solar abundance of some  minor element is a safe procedure,
and have avoided doing so in the calculations reported here.

Invalidation of the closed-box assumption we have adopted here could result
in significant changes to our conclusions. For most circumstances
the presence of inflow of unprocessed material would increase the elemental
abundances at a given redshift. An outflow of processed material will
have a smaller effect. For the probes of dust contamination, the
length (in $z$ space) of the  deuterium plateau could be strongly influenced
by inflow. Also, since in principle the $z$ dependence of any inflow/outflow
would be unknown, one could think of circumstances in which the peak of
 the supernova rates were moved to new values of $z_p$, thereby affecting
our discussions of the gamma-ray and MeV neutrino backgrounds. The assumption
of a closed box model is therefore important to our proposed tests
of dust contamination.


In summary,
assuming a closed box model
we have explored the chemical evolution of the cool neutral gas in the
early universe
using the   redshift evolution of this gas as inferred from observations
of damped Ly$\alpha$ systems. Our principal results are:- (i)
Assuming the  accuracy of the stellar yields employed here, the
abundance of the main metallicity tracer of such systems, Zn, is
significantly affected by the constant yield assumption. Direct
integration of the chemical evolution equations with redshift dependent
yields are recommended for this important element. (ii)
 Use of the O/Fe ratio can be used
to probe cosmology, although  the
 accuracy required by future observations will  make this
a difficult experiment. (iii) Other possible
signals of the
 importance of dust obscuration could be evident in  the
  deuterium  redshift
evolution, the extragalactic gamma-ray background, and the
 MeV neutrino background.
Searching for these signals will also require very high sensitivity.
 (iv) Correlations
of column density with metallicity can substantially reduce the
importance of dust related effects.

The next few years should see an ever increasing wealth of data on
the evolution of the neutral gas and elemental abundances in the
damped Ly$\alpha$ systems. Eventually this data should be able to
tightly constrain the  parameters which influence
the cosmic chemical evolution of neutral gas, such as the
 IMF and the contribution
of observational bias due to dust. This new data combined with
further theoretical study should
provide an exciting tool for probing the evolution of galaxies
in early universe.

\medskip
\noindent
The authors  would like to
acknowledge useful discussions with Dick Bond.

\vfil\eject

\centerline{REFERENCES}
\hfil\break
Bisnovatyi-Kogan, G. S. \& Seidov, Z. F., 1984,
Sov. Astron., 26, 132.
\hfil\break
Dermer, C. D.,  1986, A\&A, 157, 223.
\hfil\break
Fall, S. M. \& Pei, Y. C., 1993, ApJ, 402, 479.
\hfil\break
Gehrels, N. \& Young, C., 1995 in {\it  Extragalactic Background Radiation},
p15,
ed. D Calzetti, M. Livio \& P. Madau (Cambridge University Press: Cambridge).
\hfil\break
Gregio, L. \& Renzini , A., 1983, A\&A, 118, 217.
\hfil\break
Hamann, F. \&  Ferland, G., 1993, ApJ, 418, 11.
\hfil\break
Iben, I. \& Tutukov, 1984, ApJ Supp.,  54, 335.
\hfil\break
Krauss, L. M., Glashow, S. L. \& Schramm, D. N., 1984, Nature, 310, 191.
\hfil\break
Lanzetta, M. K., Wolfe, A. M. and Turnshek, D. A., 1995, ApJ,  440,
435.
\hfil\break
Pei, Y. C. \& Fall, S. M., 1995 ApJ, in press.
\hfil\break
Pettini, M., Smith, L. J., Hunstead, R. W. \& King, D. L., 1994, ApJ,
 426, 79.
\hfil\break
Renzini, A. \& Violi, M. 1981, A\&A, 94, 175.
\hfil\break
Silk, J. \& Schramm, D. N., 1992, ApJ Lett., 339, L9.
\hfil\break
Storrie-Lombardi, L. J., McMahon, R. G., Irwin, M. J. \& Hazard, C., 1995,
{\it ESO Workshop on QSO Absorption Lines}, ed G. Meylan,
(Berlin: Springer-Verlag).
\hfil\break
Scalo, J. M., 1986,  Fund. Cosm. Phys., 11, 1.
\hfil\break
Sneden, C., Gratton, R. G., \& Crocker, D. A., 1991, A\&A, 246, 354.
\hfil\break
Thielemann, F.K, Nomoto, K. \& Yokoi, K.  1986, A\&A, 158, 17.
\hfil\break
Thompson, I J. \& Fichtel, C. E., 1982, A\&A, 109, 352.
\hfil\break
Timmes, F.X., Lauroesch, J T. \& Truran, J. W., 1995a, ApJ, 451, 468.
\hfil\break
Timmes, F.X., Woosley, S.E.~\& Weaver, T.A.~1995b, ApJS, 98, 617.
\hfil\break
Tinsley, B. M., 1980, Fund. Cosm. Phys., 5, 287.
\hfil\break
Totani, T., Sato, K. \& Yoshii, Y., 1995, astro-ph/9509130.
\hfil\break
Vandenbergh, S. 1991, {\it Supernovae},  p711, ed. S. E. Woosley (New York:
Springer-Verlag).
\hfil\break
Weber, W. R., Simpson, G. A. \& Cane, H. V., 1980, ApJ., 263 448.
\hfil\break
Woosley S.E. \& Weaver, T.A., 1995, ApJS, in press.
\hfil\break
Woosley S.E., Wilson, J. R. \& Mayle, R., 1986, ApJ, 302, 19.

\vfil\eject

\medskip
\centerline{FIGURE CAPTIONS}

\medskip
\noindent
Figure 1. Possible evolution  histories
of $h\Omega_{\rm HI}$ as a function of redshift
($H_o=h100$ km s$^{-1}$  Mpc$^{-1}$). The solid curve  is a good match to
the recent data of Lanzetta {\it et al.} (1995); the dashed curve is a
dust-corrected representation of the solid curve; and the dotted curve
matches the recent data of Storrie-Lombardi {\it et al.} (1995)
($q_o=0.5$ assumed).
\medskip
\noindent
Figure 2. Star formation rates as a function of redshift in units
 of $\Omega_g$ per  Gyr ($q_o=0.5$ assumed).
\medskip
\noindent
Figure 3a. Redshift
 evolution of the  elemental mass fractions  (shown as log)
  for an IMF slope $x=1.7$
($q_o=0.5$ assumed).
\medskip
\noindent
Figure 3b. Redshift
 evolution of elemental mass fractions  (shown as log) for an IMF slope
$x=1.35$
($q_o=0.5$ assumed).
\medskip
\noindent
Figure 4. Redshift evolution of the O/Fe ratio (by mass)
for different cosmologies and IMF's.
\medskip
\noindent
Figure 5. Redshift
 evolution of elemental  mass fractions  (shown as log)
for an IMF slope $x=1.7$ and
dust correction included.
($q_o=0.5$ assumed).
\medskip
\noindent
Figure 6. Redshift evolution of the ratio of the
 deuterium abundance to its primordial abundance ($q_o=0.5$ assumed).
\medskip
\noindent
Figure 7. Type~II supernova rates shown here as $R_{II}/\rho_c$.
($R_{II}$ in units Gyr$^{-1}$  per unit comoving volume).
\medskip
\noindent
Figure 8. Effects of dust on $f(N,z)$ for constant and variable
gas-to-dust ratios.
\medskip
\noindent

\vfil
\eject

\centerline{APPENDIX}
\medskip
\noindent
For reference, the numerical output of the mass fractions for our
calculations assuming the solid curve of figure~1, $q_o=0.5$,
 and IMF slope of
$x=1.7$, is shown in table~1.
\vskip 1.0truein

\tolerance = 1500
\baselineskip = 12pt

\def\gsim{\raise.2ex\hbox{$>$}\kern-.75em \lower.7ex\hbox{$\sim$}}
\def\lsim{\raise.2ex\hbox{$<$}\kern-.75em \lower.7ex\hbox{$\sim$}}
\hoffset = .7cm
\hsize=6.0truein
\centerline{TABLE 1.}
\vskip .1cm
\vbox{
\centerline{Element Mass Fractions  (-log)}
\vskip.1truein
\hrule width 6.0truein
\vskip.03truein
\hrule width 6.0truein
\vskip.1truein
 \halign{\hskip10pt#\hfil \tabskip.4truein & #\hfil \tabskip.4truein & #\hfil
\tabskip.4truein & #\hfil \tabskip.4truein & #\hfil \tabskip.4truein & #\hfil
\tabskip.4truein & #\hfil \tabskip.4truein & #\hfil \tabskip.4truein & #\hfil
\tabskip.4truein & #\hfil
\tabskip.5truein & #\hfil\cr
\hfil z & \hfil C & \hfil N  & \hfil O & \hfil Mg & \hfil Si &
 \hfil Ca & \hfil Cr & \hfil Fe & \hfil Zn \cr
        \openup1pt\cr
        \noalign{\hrule width 6.0truein}
        \openup3pt\cr

    5.0   &4.8  & 5.3  & 3.9  & 5.3  &  4.9   & 6.1 &   6.7  &  4.9  &  8.2\cr

    4.5  & 4.6  & 4.9 &  3.7  & 5.0  &  4.6  &  5.8   & 6.5   & 4.7  &  7.9\cr

    4.0 &  4.3  & 4.6 &  3.4 &  4.8  &  4.4  &  5.6 &   6.2   & 4.4 &   7.7\cr

    3.5  & 4.1 &  4.3  & 3.2  & 4.6  &  4.2   & 5.4   & 6.0 &   4.2  &  7.5\cr

    3.0 &  3.8 &  4.1 &  3.1 &  4.4  &  4.0  &  5.2  &  5.8   & 4.0   & 7.3\cr

    2.5 &  3.6  & 3.9 &  2.9 &  4.2  &  3.8  &  5.0 &   5.6  &  3.7  &  7.1\cr

    2.0 &  3.3  & 3.7  & 2.7  & 4.1 &   3.7  &  4.8  &  5.4 &   3.5  &  6.9\cr

    1.5 &  3.1 &  3.5 &  2.6 &  3.9 &   3.5  &  4.7 &   5.2  &  3.3  &  6.7\cr

    1.0 &  3.0  & 3.4  & 2.5 &  3.8  &  3.4   & 4.5  &  5.0  &  3.0  &  6.5\cr

    0.5  & 2.9 &  3.3 &  2.4  & 3.6  &  3.2  &  4.3  &  4.8  &  2.8  &  6.3\cr
 }

        \vskip.1truein
        \hrule width 6.0truein
        }
\vfill\eject
\bye